\documentclass[twocolumn,showpacs,preprintnumbers,amsmath,amssymb]{revtex4}
\usepackage{graphicx}
\usepackage{dcolumn}
\usepackage{bm}

\begin{document}
\title{ Graphene  nanosystems  and  low-dimensional Chern-Simons topological insulators}
\author{Y. H. Jeong and S. -R. Eric Yang
\footnote{corresponding author, eyang812@gmail.com}  }
\affiliation{  Department of Physics, Korea  University, Seoul,
Korea\\}

\begin{abstract}
A graphene  nanoribbon is a  good candidate for  a $(1+1)$ Chern-Simons topological insulator since it  obeys particle-hole symmetry.
We show that in a finite  semiconducting  armchair ribbon,  which has  two zigzag edges and two armchair edges, a $(1+1)$ Chern-Simons topological insulator is indeed realized as the  length of the armchair edges   becomes large in comparison to that of the zigzag edges.
But only a quasi-topological insulator is formed in a  metallic armchair ribbon with a pseudogap.  In such  systems a zigzag edge acts like  a domain wall, through which  the polarization changes from $0$ to $e/2$, forming  a fractional charge of one-half.    When  the lengths of
the zigzag edges and  the armchair edges are comparable  a  rectangular graphene sheet (RGS) is realized, which also possess  particle-hole symmetry.  We show that it is a $(0+1)$ Chern-Simons topological insulator.
We find that the  cyclic Berry phase of  states of  a RGS is quantized as  $\pi$ or $0$ (mod $2\pi$), and that the Berry phases of the particle-hole conjugate states are equal each other.  By applying  the Atiyah-Singer index theorem to a  rectangular ribbon and a RGS we find that the lower bound on the
number of nearly zero energy  end states  is approximately proportional  to the length of the zigzag edges.
However, there is  a
correction to this index theorem due to the effects  beyond the effective mass approximation.
\end{abstract}

\maketitle

\section{Introduction}

Recently a rapid progress\cite{Novo,Zha,Cai,Kato} has been made in the
fabrication of   graphene nanoribbons\cite{Fuj,Neto,Brey,Son,Yang,Lee} .
They have a great potential for spintronic applications, where
electronic many-body  interactions may play a significant
role\cite{Son,Yang,Lee}.  Nanoribbons may also provide numerous
interesting issues in fundamental physics, such as topological insulators\cite{Bern,topoins1,topoins2,Kit,Hee,Wen,Zhang,Jeong}.

A nanoribbon is  an excellent candidate for a  Chern-Simons topological insulator since it obeys particle-hole symmetry,  which is one of the
symmetry requirements of  low dimensional Chern-Simons topological insulators
(this symmetry plays an important role in the $Z_2$ classification of the Hamiltonians).
A $(1+1)$ dimensional Chern-Simons topological insulator is defined by  a Lagrangian density\cite{Zhang}
which involves the electric polarization $\mathcal{P}$\cite{Van,Res,Zak}
\begin{eqnarray}
\mathcal{L}=\mathcal{P}\epsilon^{\mu\nu}\partial_{\mu}A_{\nu},
\label{Lang1}
\end{eqnarray}
where  $A_{\nu}$ is the electromagnetic field.
From   the Lagrangian density  follows that
the current density  is related to the polarization
\begin{eqnarray}
j_{\mu}=-\epsilon_{\mu\nu}\frac{\partial \mathcal{P}}{\partial y_{\nu}}.
\label{jvsP}
\end{eqnarray}
According to this Lagrangian  one
needs to show  the presence of  electric polarization to establish that  a nanoribbon  is indeed a Chern-Simons topological insulator.  A  $(0+1)$ dimensional topological insulator can be also defined by a Chern-Simons
effective Lagrangian\cite{Zhang}
\begin{eqnarray}
\mathcal{L}=Tr[B_0],
\label{Lang0}
\end{eqnarray}
where $B_0$ is a  Berry vector potential and
the trace stands for  the sum over the occupied states.

In the Chern-Simons theory   one-dimensional Hamiltonians are labeled\cite{Zhang} by the parameter $\theta$ such that for $\theta=0$ the Hamiltonian is $Z_2$ trivial and for    $\theta=\pi$
the Hamiltonian is $Z_2$ non-trivial.  As  the parameter $\theta$ changes by  $2\pi$,  the polarization change, $\Delta \mathcal{P}=\mathcal{P}(2\pi)-\mathcal{P}(0)$, is given by
the surface integral over a toroidal surface with the value equal to an integer in units of $e$  (a first Chern number).
Here the polarization of an occupied band is defined by the Zak phase $Z$,   obtained   from  { \it dimensional reduction} by performing a
one-dimensional cut of a two-dimensional Brillouin zone\cite{Del}\begin{eqnarray}
 \mathcal{P}(\theta)=\frac{e}{2\pi}Z=\frac{e}{2\pi}\oint dk_y (-i)\left\langle \theta,k_y \right|\frac{\partial}{\partial k_y}\left|\theta,k_y\right\rangle,
 \label{Zak}
\end{eqnarray}
where  the $\left|\theta,k_y\right\rangle$  are the periodic part of two-dimensional Bloch wavefunctions
(from now one we will call this quantity the {\it reduced} Zak phase to contrast it to the one-dimensional Zak phase\cite{Zak} defined in Eqs. (\ref{pol}) and (\ref{vecp})).
When the integration over $k_y$ in the reduced Zak phase is performed  along a zigzag edge the Zak phase is zero while when integrated along an armchair edge it is $\pi$\cite{Del}.
In contrast, it should be noted that,  in translationally invariant one-dimensional systems  the ordinary definition of the Zak phase is
a multivalued quantity with a quantum uncertainty\cite{Zak,topoins2,Hat} (we will show this explicitly for graphene nanoribons).

When translational invariance is broken the spatial variation of polarization  can change  abruptly  and domain walls can be present in graphene nanoribbons:  in a  graphene zigzag ribbon translational invariance is broken along the {\it perpendicular} direction  to  the zigzag edges and
the polarization changes suddenly across the zigzag edges.  This effect is  described by the Chern-Simons field theoretical result Eq.(\ref{jvsP}), where the coordinate $y$ is along   the perpendicular direction. The domain wall has  nearly zero energy zigzag {\it edge} states,  similar to quantum Hall  systems.   A periodic  armchair ribbon is translationally  invariant  along
the ribbon direction.   This invariance can be broken by  cutting the bonds transversely, which will produce two extra zigzag edges (see Fig.\ref{rect}).
This will  gives rise to the spatial variation of polarization along the ribbon direction, and   produce {\it end} states (domain walls) with nearly zero energy\cite{Kit}.  In such finite graphene armchair ribbons\cite{Fuj,Yang,Brey,Lee}
we find that  a $(1+1)$ Chern-Simons topological insulator is realized as the aspect ratio between the lengths of armchair and zigzag edges
goes $L_y/L_x\rightarrow \infty$.     We will show from the relation between charge and polarization, Eq.(\ref{jvsP}),
that  end states have a fractional charge, analogous to    polyacetylene\cite{Hee}: the polarization $P(y)$ varies from $0$ to $e/2$ as a function of $y$ along the ribbon direction.  This is an important implication of the  Lagrangian density, and  is
a hallmark of a non-trivial topological insulator.
The corresponding charge density of an end state  for metallic and semiconducting armchair ribbon   decays   exponentially with  very different
the decay lengths.   In the metallic case with a pseudogap  the decay length is comparable to the system size, and only a {\it quasi-topological} insulator is formed.

When the aspect ratio is $L_y/L_x\sim 1$  under the condition $L_x \lesssim 100\AA$ a RGS\cite{Tang,Kim0} is formed.  We show that it is a $(0+1)$ Chern-Simons topological insulator with particle-hole symmetry intact.
In this island-like  system  a Berry phase can be generated  through modulation of the hopping parameter of a C-C
bond.   We find that some states of  a RGS have a  cyclic Berry phase of $\pi$ mod $2\pi$
while other states have zero Berry phase.   In addition,  the Berry phases  of particle-hole conjugate states  are equal to each other mod $2\pi$.
We will show that it
has also nearly zero energy edge states, which  is another hallmark of topological insulators.

In both $(1+1)$ and $(0+1)$ Chern-Simons topological insulators we find, using the Atiyah-Singer index theorem\cite{atiyah},
that number of nearly zero energy end states   is proportional to the length of the zigzag edges.
We find also a  correction to this Atiyiah-Singer result  due to the effects  beyond the effective mass approximation.

This paper is organized as follows.   We compute the polarization near a domain wall  in a finite graphene ribbon In Sec.II.
In Sec.III, we compute  a  Berry phase of  a RGS, and   show that such a system is a  $(0+1)$ dimensional topological Chern-Simons insulator.
The number of zero modes of a finite length zigzag edge  is computed in Sec.IV, in addition to a  correction to the Atiyah-Singer index theorem on the lower bound on the number of zero modes.  Conclusions are given in Sec. V.   In Appendix    we show that the polarization a one-dimensional  nanoribbon  has a quantum uncertainty between the values  $0$ and  $\frac{e}{2}$ (modulo $e$).

\section{Armchair ribbon with broken translational invariance: domain wall and  fractional charge }

\begin{figure}[!hbpt]
\begin{center}
\includegraphics[width=0.25\textwidth]{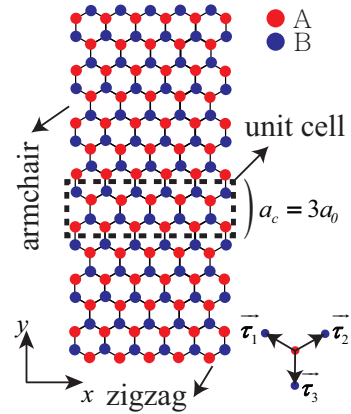}
\end{center}
\caption{  Finite graphene  ribbon has
two zigzag edges and two armchair edges.
When the length of the zigzag  edges is  $L_x=3La_0$ or $(3L+1)a_0$
a gap is present, but for $L_x=(3L+2)a_0$ no gap exists ($L$ is an integer and $a_0=\sqrt{2}a$ is the unit cell length of the honeycomb lattice).   Vectors $\vec{\tau}_l$ connect  a A carbon atom to  three neighboring B carbon atoms. A unit cell of an armchair ribbon is shown in the dashed box. }
\label{rect}
\end{figure}

\begin{figure}[!hbpt]
\begin{center}
\includegraphics[width=0.3\textwidth]{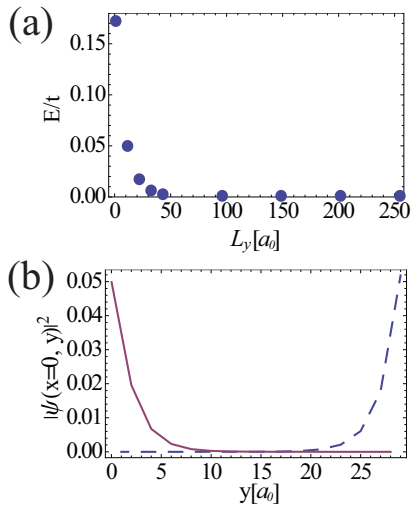}
\caption{Results for finite  semiconducting armchair ribbon  with a short  width equal to the length of zigzag edges $L_x=3La_0=3a_0$.  (a) The eigenenergy of an end  state  vs the length armchair edges   $L_{y}$.  (b) Plot of the probability density, $|\psi_M(x=0,y)|^2$, of the state shown in (a)   vs  $y$ for the length armchair edges  $L_y=30a_0$.    The left zigzag edge ($y=0$) consists of $A$-carbons and the right edge ($y=3.52a_0$) of $B$-carbons. The red  line indicates the value of the probability density on $A$-carbons and the blue  line indicates the probability density on $B$-carbons.   }
\label{rect-anal1}
\end{center}
\end{figure}

\begin{figure}[!hbpt]
\begin{center}
\includegraphics[width=0.3\textwidth]{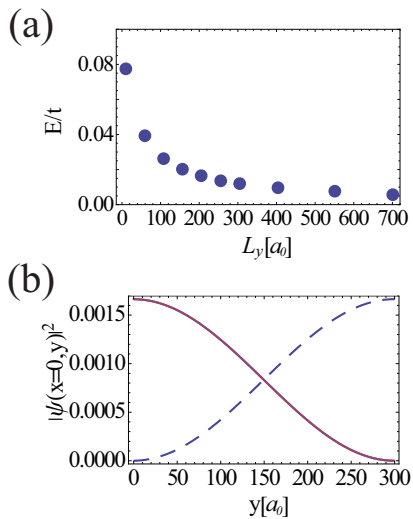}
\caption{Same as in Fig.\ref{rect-anal1} but for  the metallic case with the length zigzag edges  $L_x=(3L+2)a_0=2a_0$.    From (b) we see that the  width of domain wall is long, comparable to the length of the system $ L_y=300a_0$.    }
\label{rect-anal2}
\end{center}
\end{figure}

Here we consider a spatial variation of the polarization and the formation of a fractional charge state.
Cutting transversely the bonds of a periodic an armchair ribbon  breaks translational invariance and generates  a {\it finite} length one-dimensional system with two zigzag edges at ends of the system, see Fig.\ref{rect}.   Then outside of the armchair the system is  trivial  with $\theta=0$, while  inside the ribbon  the system  is  non-trivial
with $\theta=\pi$.   In such an armchair ribbon, as the length of armchair edges   becomes longer,  doubly degenerate end  states with nearly zero energy appear, as our tight-binding numerical results  shown  in  Fig.\ref{rect-anal1}(a) demonstrate (the magnetic flux $\phi$ is set to zero).  Here the length of zigzag edges is  $L_x=3La_0$ so that an energy gap\cite{Lee} exist.  In the limit where the length of the armchair edges $L_y\rightarrow \infty$ we find that the decay length of the probability density of the end state is short, comparable to  $a_0$, see Fig.\ref{rect-anal1}(b).
This  behavior is typical of a zigzag edge state\cite{Neto}.
These results are  also true for
a semiconducting armchair ribbon with  different  length of zigzag edges $L_x=(3L+1)a_0$.
In the metallic armchair ribbon with   the length zigzag edges
$L_x=(3L+2)a_0$ nearly zero energy
end  states also appear, see Fig.\ref{rect-anal2}(a).   However, as shown in   Fig.\ref{rect-anal2}(b) the decay length of these states are much longer, comparable to the system length.   We will thus call this metallic armchair ribbon with a pseudogap a {\it quasi-topological insulator}.

These  probability densities satisfy the relation between polarization and charge of a Chern-Simons topological insulator, Eq.(\ref{jvsP}):
for a semiconducting  armchair ribbon (with a  gap in the DOS) the A- or B-type probability density decays rapidly
\begin{eqnarray}
\rho(y)&=&\frac{\partial \mathcal{P}(y)}{\partial y}\sim e^{-\alpha\frac{y}{ a_c}}.\nonumber\\
\end{eqnarray}
  For a metallic armchair ribbon  (with a  pseudogap in the DOS) it decays slowly
\begin{eqnarray}
\rho(y)&=&\frac{\partial \mathcal{P}(y)}{\partial y}\sim e^{-\beta\frac{y}{ L_y}},\nonumber\\
\end{eqnarray}
where $L_y$ is the length of the armchair   edges.    In the limit $L_y\rightarrow \infty$ the integrated probability density on one type of carbon atoms is
gives the fractional charge, given by\cite{Zhang}
\begin{eqnarray}
Q&=&\int _{0}^{\infty}dy \frac{\partial \mathcal{P}(y)}{\partial y}=P(\infty)-P(0)=-\frac{e}{2}.
\label{frac}
\end{eqnarray}
The existence of this charge of one-half is thus related to the variation the polarization $P(y)$ from $0$ to $-e/2$.
These end states with zero energy represent a fractional charge of $1/2$\cite{Kit,Hee,Jeong}
of a domain wall  (they appear in degenerate pairs\cite{Jeong,Kane0}).

\section{Rectangular sheet: $(0+1)$ dimensional  Chern-Simons topological insulator }

\begin{figure}[!hbpt]
\begin{center}
\includegraphics[width=0.3\textwidth]{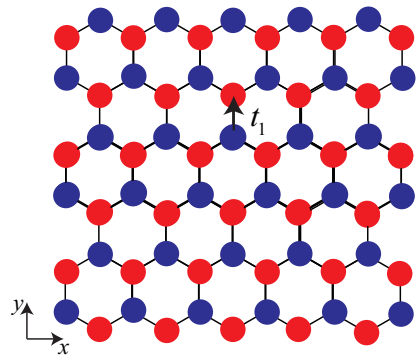}
\caption{ RGS is shown.  The   length of armchair (zigzag) edges is $L_y (L_x)$.  We choose  a C atom and  distort  its hopping integral   $t_1$.  Adiabatic cycle is performed through the variation of $t_1$.    }
\label{recsh}
\end{center}
\end{figure}

In the opposite regime  $L_y/L_x\sim 1$  with $L_x\lesssim 100\AA$ a RGS is realized, see  Fig.\ref{recsh}.  A RGS is an island-like system and is   a $(0+1)$ dimensional system.
However,  the effects of
armchair and zigzag edges may compete.
It is thus unclear whether a RGS is a topological insulator.  Here we investigate whether a RGS is really a Chern-Simons topological insulator.
To establish that its effective Lagrangian is given by  a $(0+1)$ Chern-Simons topological field theory, Eq.(\ref{Lang0}),
we need to construct non-vanishing Berry vector potentials.

It is not trivial to choose the appropriate adiabatic cyclic parameters.  We use the following adiabatic cyclic parameters.
We  induce a time-dependent
change of a C-C bond, see  Fig.\ref{recsh}.  Its hopping parameter has time dependence  $t_1=0.9te^{i2\pi \tau/\tau_0}$ with $0\leq \tau/\tau_0 \leq1$.
Note that this perturbation preserves particle-hole symmetry.
The Berry vector potential  of  this $(0+1)$ dimensional system  is
\begin{eqnarray}
B_0=\sum_{\alpha }'  \left\langle\psi_{\alpha}  \right|i\frac{d}{d \tau}\left|\psi_{_{\alpha}}  \right\rangle,
\end{eqnarray}
where the summation is over the occupied states.  We use the tight-binding method in site-representation to compute the Berry phase since tight-binding approach is more accurate than the effective mass approach for small island-like systems.
We use a gauge fixing to ensure numerically stable results\cite{Hat}.

\begin{figure}[!hbpt]
\begin{center}
\includegraphics[width=0.35\textwidth]{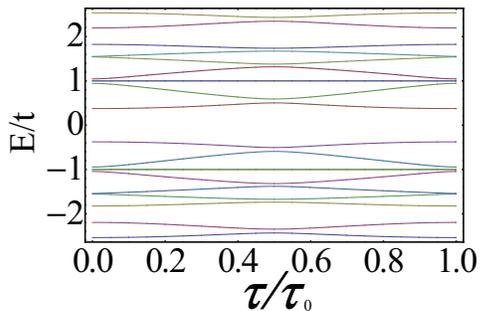}
\caption{ Energy levels of a RGS vs $\tau$. The lengths are $(L_x,L_y)=(3a_0,3.52a_0)$. Particle-hole symmetry is present.   }
\label{energy}
\end{center}
\end{figure}

\begin{figure}[!hbpt]
\begin{center}
\includegraphics[width=0.3\textwidth]{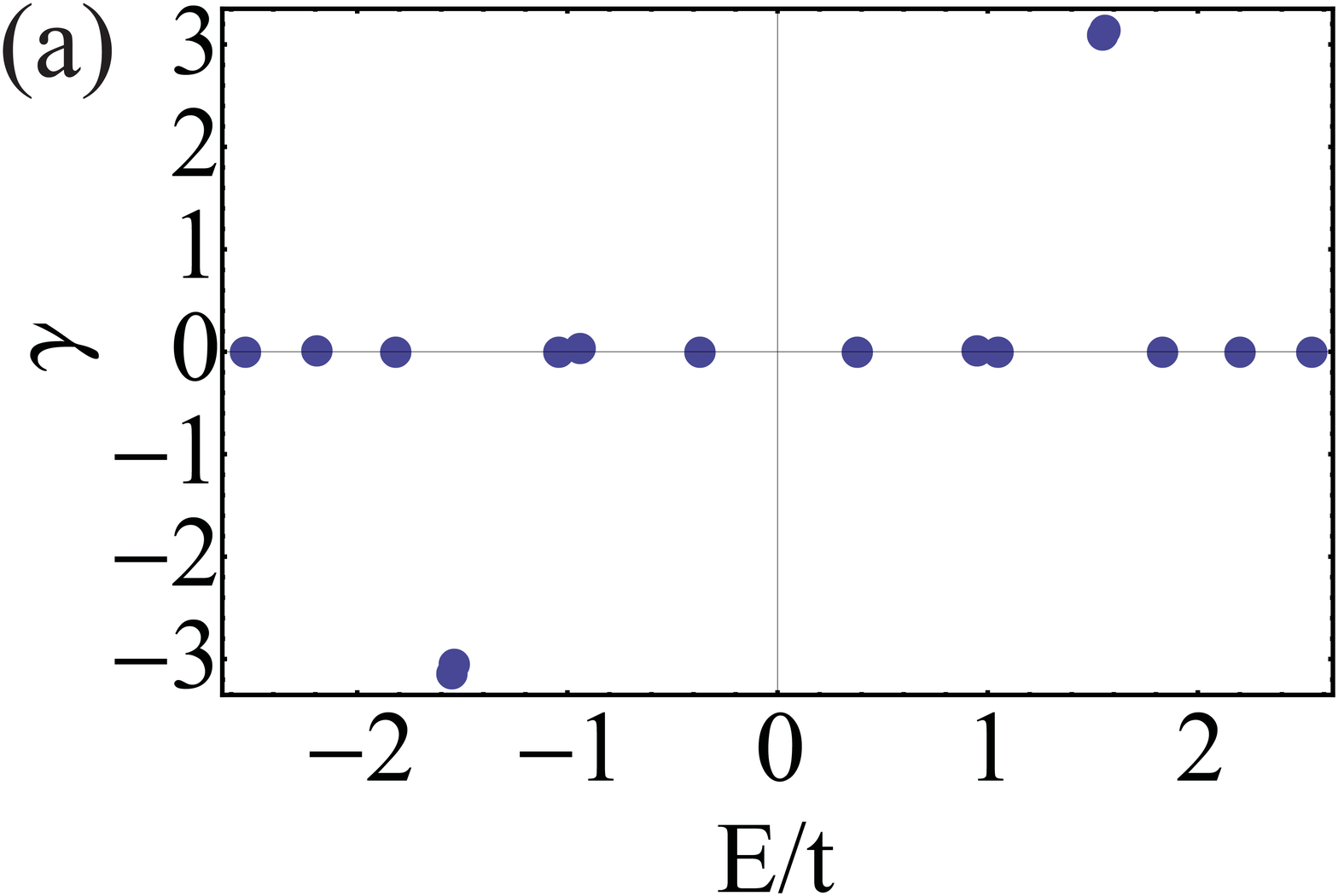}
\includegraphics[width=0.3\textwidth]{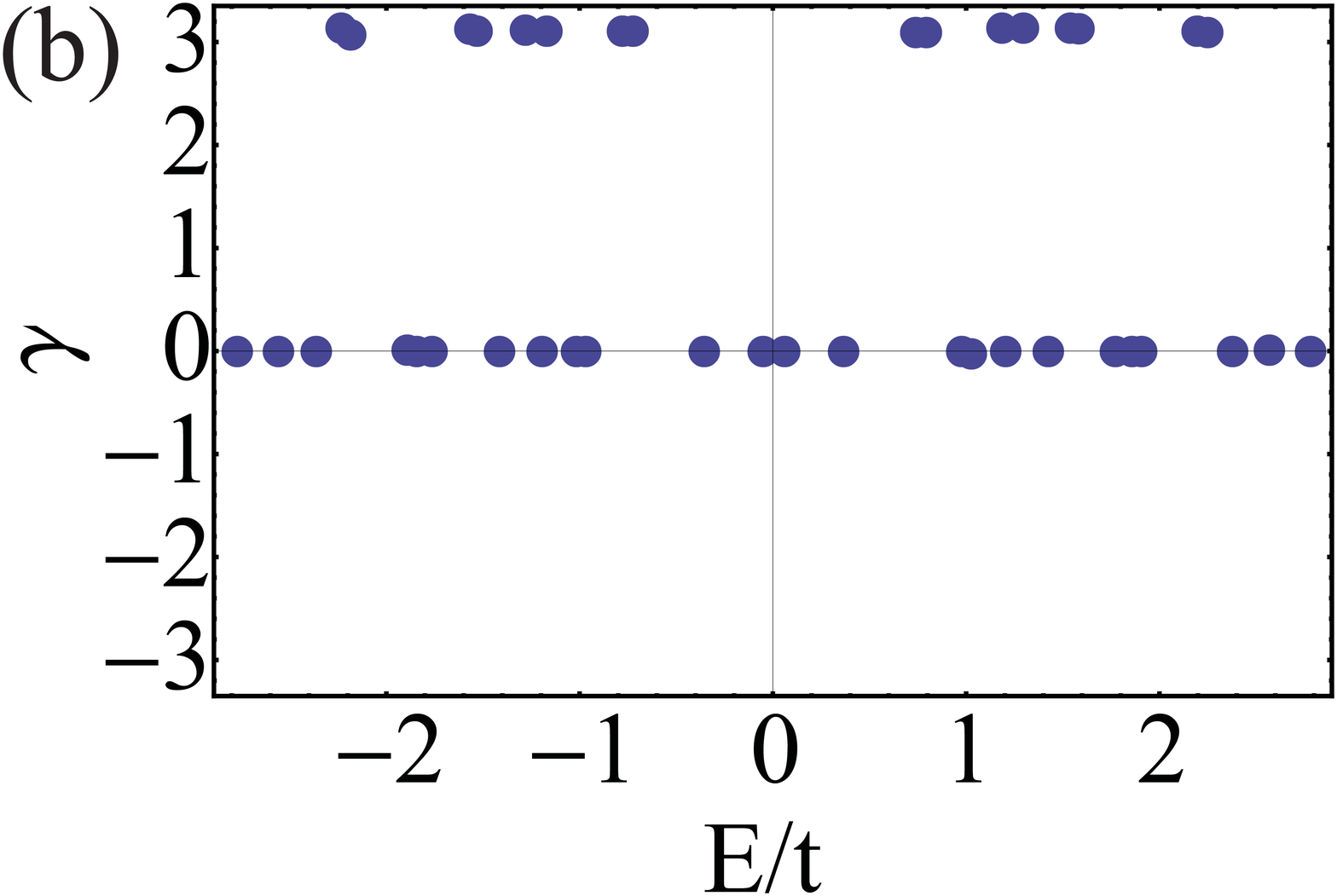}
\caption{Berry Phase of a RGS vs eigenenergies $E$ when  $(L_x,L_y)=(3a_0,3.52a_0)$ (a) and   $(L_x,L_y)=(5a_0,3.52a_0)$ (b).
    In (a) we exclude points at $E=\pm t$ because they are degenerated.}
\label{Berry}
\end{center}
\end{figure}

To compute the Berry phase we first need to obtain eigenstates and eigenenergies as a function of the adiabatic parameter.  The energy levels of a RGS   are shown in Fig.\ref{energy} as a functions of time $\tau$.
Note that the computed energies display  particle-hole symmetry.
The computed Berry phases of the levels  are
either  zero or $\pi$ mod $2\pi$\cite{Hat}, see Fig.{\ref{Berry}}(a) ($L_x=3La_0$).  Note the Berry phases of a pair of particle-hole conjugate states is equal each other mod $2\pi$.  We have also found  that the nearly zero energy zigzag edge states exist, see Fig.\ref{energy}.  It has aslo nearly zero energy edge states, which  is another hallmark of topological insulators,

When the length zigzag  edges  is
$L_x=(3L+2)a_0$ the Berry phase is  again either  zero or $\pi$ mod $2\pi$, and
the states  in a particle-hole conjugate have the  same value of Berry phase mod $2\pi$.

\section{Number of zero modes}

So far  we have investigated nearly zero modes of narrow ribbons, i.e, zero modes of one-dimensional systems.   We now study quasi-one-dimensional ribbons that have broader widths.
We  compute, as a function of the ribbon width,  the number of  nearly zero energy edge modes\cite{Del,Ryu} a finite length zigzag edge supports.  The localization length of these
edge modes increases with increasing energy.
We choose to employ
the Atiyah-Singer index theorem\cite{Pac,atiyah} to compute the number of edge modes.   We will show that it
is proportional to the length of the zigzag edges.  There is   also a correction to this result, which we compute.

A RGS can be made out of  an armchair ribbon by cutting C-C bonds along the transverse direction of the ribbon, which produces two zigzag edges.
Mathematically this can be done by applying  a chiral vector potential $\vec{A}_c=(A_{c,x}\Theta(y),0,0))$ along the x-axis  in a  rectangle  \cite{Jeong} (this area can be chosen as the unit cell shown in Fig.\ref{rect}).  The bonds will be cut at the critical strength
\begin{eqnarray}
\frac{eA_cv_F}{c}=t,
\end{eqnarray}
where $\hbar v_F=\frac{3}{2}t  a$ with the C-C distance $a$.
Since the vector potential is chiral the relevant $\mathbf{K}$ and $\mathbf{K'}$ Dirac Hamiltonians are:
\begin{eqnarray}
H_{\mathbf{K}}=v_{F}\vec{\sigma}\cdot
(\vec{p}-\frac{e}{c}\vec{A}_c(\vec{r}))
\end{eqnarray}
for the $\mathbf{K}$ valley, and
\begin{eqnarray}
H_{\mathbf{K'}}=v_{F}\vec{\sigma}'\cdot
(\vec{p}+\frac{e}{c}\vec{A}_c(\vec{r}))
\end{eqnarray}
for the $\mathbf{K'}$ valley ($e>0$).
The x and y components of
$\vec{\sigma}'$ are $-\sigma_{x}$ and $\sigma_{y}$. Note that,
unlike for the real vector potential, the chiral vector potential
appears with opposite signs for $\mathbf{K}$ and
$\mathbf{K'}$ valleys  (it is chiral in the sense that it
distinguishes valleys).
In a RGS $\mathbf{K}$ and    $\mathbf{K'}$ valleys are coupled by the armchair edges\cite{hclee}, and the effective mass Hamiltonian of a RGS is
a block diagonal  matrix with
$H_\mathbf{K}$   and  $H_\mathbf{K'}$ as blocks.

\begin{figure}[!hbpt]
 \begin{center}
 \includegraphics[width=0.35\textwidth]{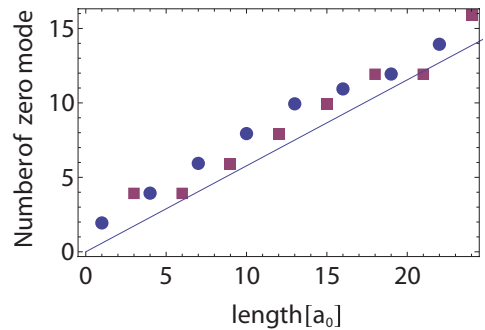}
 \end{center}
 \caption{Number of zero modes of a RGS, computed using the tight-binding Hamiltonian, are plotted
 as a function of the length   of the zigzag edges $L_x/a_0$.
 Here the length of the armchair edges is $L_y=143.6a_0$. Squares are for $L_x=3La_0$ and circles are for $L_x=(3L+1)a_0$.  Note the number does not increase linearly with $L_x$.   The lowest curve represents the lower bound and is given by  the Atiyah-Singer result of Eq.(\ref{AS}).}
 \label{number}
 \end{figure}

Since the Dirac operator is an elliptic operator, it is possible to
employ the Atiyah-singer index theorem\cite{Pac} (In the case of non-compact
surfaces the theorem is applicable only when there is no  net flux going through the open faces\cite{Pac0}).
To apply the theorem  we rewrite  the $\mathbf{K}$ valley Hamiltonian as
\begin{equation}
H_{\mathbf{K}}=\left(
\begin{array}{cc}
0 & P^{\dag}\\
P&0\\
\end{array}
\right),
\end{equation}
where the operator $P^{\dag}=\hbar v_F(\partial_{x}-i\partial_y-\frac{ie}{\hbar c}A_{c,x})$.  Then
\begin{equation}
H_{\mathbf{K}}^2=\left(
\begin{array}{cc}
P^{\dag}P & 0 \\
0 & PP^{\dag}
\end{array}
\right).
\end{equation}
We define   $\nu_{+}$ as the number of zero modes of $P^{\dag}P$ and $\nu_{-}$ as the number of zero modes of $PP^{\dag}$.  Since there is no net flux  the Atiyah-Singer index theorem dictates the  difference in the numbers of zero modes is related to the  vector potential
\begin{eqnarray}
\mathrm{index}(H_{\mathbf{K}})=\frac{1}{2\pi\phi_0}\oint_C\vec{A}_c\cdot d\vec{l}=\nu_{1}-\nu_{2}=0,\nonumber\\
\end{eqnarray}
where the contour C is along the edges of the rectangle.    Non-zero contributions come from contour pieces $C_1$   and $C_2$, where $C_1=-C_2$.
We have checked numerically that the numbers of zero modes from the contours $C_1$  and $C_2$
are identical, i.e., $\nu_{1}=\nu_{2}$.  It should be noted that these zero modes  also have  {\it identical} wavefunctions.
This implies that the index $\nu_1$ is  proportional to the line integral along $C_1$.
The proportionality constant can be determined from the comparison with the tight-binding numerical result for the number of zero modes:
we  find that  the total number of zero energy modes  is proportional to
the length   of the zigzag edges  $L_x$
\begin{eqnarray}
N_0=   \frac{1}{2\phi_0}\int_{C_1} \vec{A}_c\cdot d\vec{l}   =\frac{A_cL_x}{2\phi_0}=\frac{1}{\sqrt{3}}\frac{L_x}{a_0}.
\label{AS}
\end{eqnarray}
This number is equal to lower limit of the number of zero modes, see Fig.\ref{number}.
The tight-binding result shows that there is  a correction to this  result due to the effects beyond the effective mass approximation, see Fig.\ref{number}.

 \begin{figure}[!hbpt]
 \begin{center}
 \includegraphics[width=0.35\textwidth]{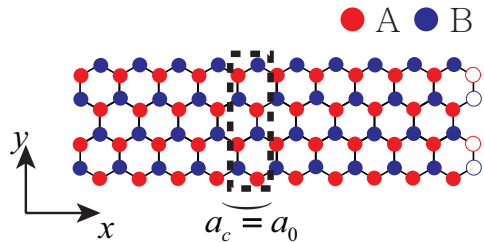}
 \end{center}
 \caption{A unit cell of a periodic zigzag ribbon is shown. The length of the unit cell is  $a_c=a_0$.}
 \label{PZGR}
 \end{figure}

\begin{figure}[!hbpt]
\begin{center}
\includegraphics[width=0.35\textwidth]{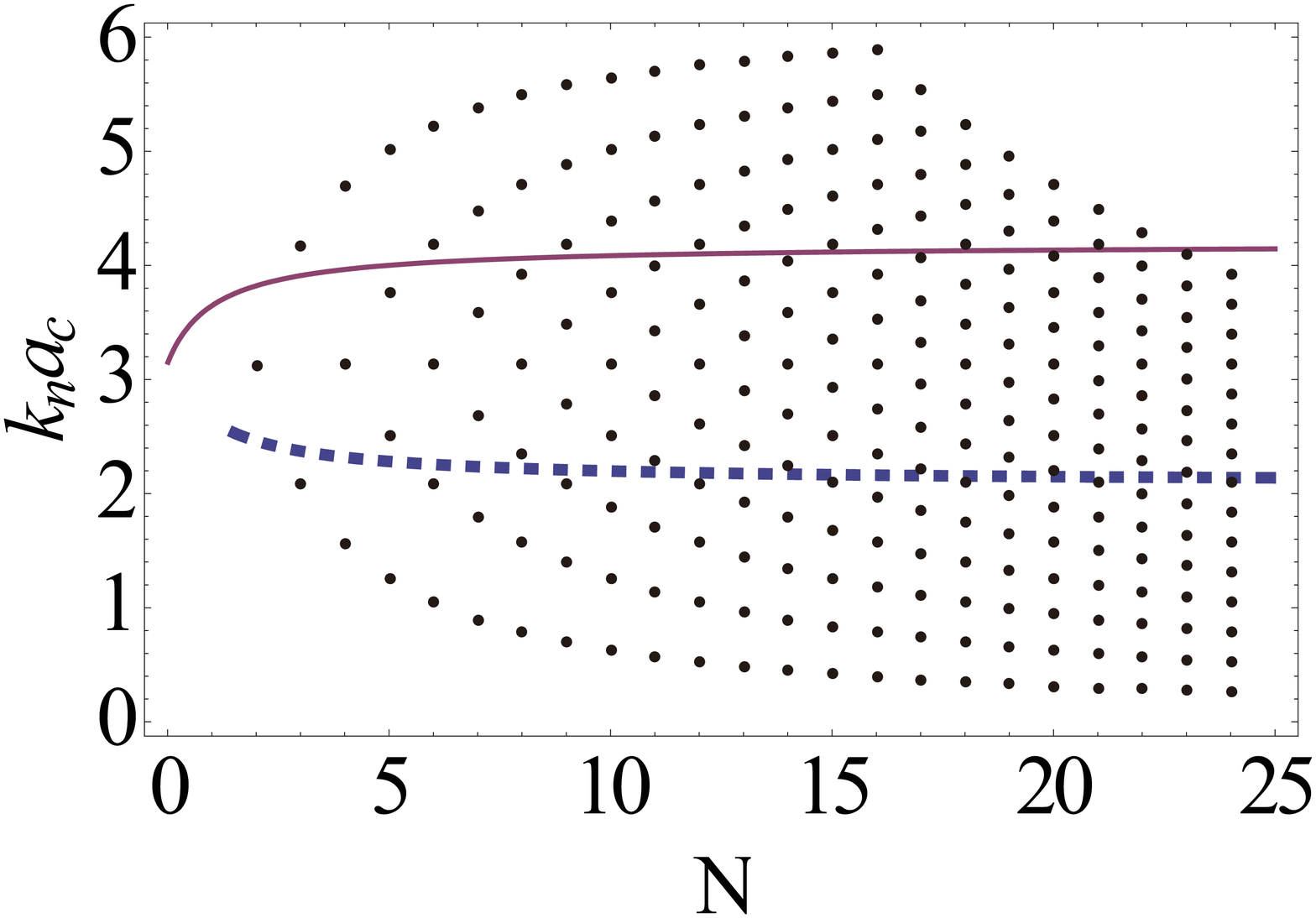}
\end{center}
\caption{ For a periodic zigzag ribbon  nearly zero modes exist at wavevectors   $k_na_0=2\pi/N, 4\pi/N, .....14\pi/N$. These values are plotted as a function of $N$.
 Solid and dashed lines represent  lower and upper bounds $k_c^-(N)$   and $k_c^+(N)$.}
\label{deviation}
\end{figure}

There is another way to compute the number of nearly zero energy modes of a  zigzag edge of length $L_x$.  Consider a periodic zigzag  ribbon with  a finite length $L_x$, see Fig.\ref{PZGR}.   For this system we can compute the exact number of zero modes using the tight-binding approach.
In the tight-binding calculation\cite{Waka,Kim1}  the nearly
zero eigenenergies are positive in the interval $k_c^+(N)<k_n<\pi/a_0$  while  they are  negative in the interval $\pi/a_0<k_n<k_c^-(N)$,
where the  values of the critical wavevectors are
\begin{eqnarray}
k_c^{\pm}(N)a_0=  2cos^{-1}\left(\pm\frac{1/2}{1+1/N}\right)
\end{eqnarray}
with the periodic  length $L_x=Na_0$ ($N$ is the number of unit cells cells and $a_0$ is the unit cell length, see Fig.\ref{PZGR}).  At the values $k_c^+(N)$ and $k_c^-(N)$ eigenenergies split from the bulk graphene
energy spectrum.  The number values of $k_n$ satisfying these inequalities $k_c^+(N)<k_n<k_c^-(N)$ are shown in   Fig.\ref{deviation}.    We find that, as $N$ increases,
this number  does not always increase monotonously, in contrast to the result of Atiyah-Singer, Eq.(\ref{AS}).  Note that solutions at each $k_n$ are nearly
degenerate due to particle-hole symmetry, which gives an additional factor of $2$.  Thus, in total,  the number of solutions
must be multiplied by 2.   This result for a zigzag ribbon is in agreement with the result of RGS for $L_x\gg a_c$:
for $N=21$ we find  $12$ zero modes in both systems,  see Figs.\ref{number} and \ref{deviation}.
Also it should be noted that the numerical values of the number of zero modes are all even, consistent with the $Z_2$
classification\cite{Kane0}.

\section{summary}

One of the
symmetry requirements of low dimensional Chern-Simons topological insulators is particle-hole symmetry.  Another important ingredient of the $(1+1)$ Chern-Simons theory is a finite electric polarization defined by the Zak phase obtained from dimensional reduction. In a finite armchair ribbon with broken translational invariance   a $(1+1)$ Chern-Simons topological insulator is realized as the aspect ratio  between the lengths of armchair and  zigzag edges   goes $L_y/L_x\rightarrow \infty$.  However,  only a quasi-topological insulator is formed in such a system when the energy gap goes to zero , i.e., in the  metallic case with a pseudogap.   In the opposite limit  $L_y/L_x \sim 1$ a $(0+1)$ Chern-Simons topological insulator is realized, and
the  cyclic Berry phase  is quantized as  $\pi$ or $0$ (mod $2\pi$).   The Berry phases of particle-hole conjugate states are equal each other.
In both   armchair ribbon and RGS the number of  nearly zero energy end modes  is proportional to the length of the zigzag edges. A correction to  this result that includes effects ignored in the effective mass approach is computed.

It will be interesting to measure the number of zero modes and the fractional charge in these systems.
When a magnetic field is applied to a RGS time reversal symmetry is broken and more zero energy states will be formed\cite{Kim0,KimAP}.

\appendix
\section{Quantum uncertainty of polarization in nanoribbons }

As we mentioned in Sec.I  the one-dimensional   Zak phase of periodic/infinite  systems is
a multivalued quantity with a quantum uncertainty (this is {\it not} the reduced Zak phase that enters in the Chern-Simons theory).
When inversion/particle-hole  symmetry is present in such a system polarization is quantized and can only take the value $0$ or $\frac{e}{2}$ (modulo $e$)\cite{topoins2,Zhang,Zak,Hat}.
Although the unit cell of a quasi-one-dimensional   ribbon can contain numerous carbon atoms (see Fig.\ref{rect}) its polarization  is expected to be quantized with the value $0$ or $\frac{e}{2}$ (modulo $e$).  Here we  show  this  explicitly  using a gauge invariant method.

We compute the one-dimensional Zak phase   and polarization  in the presence of a time-dependent
vector potential $A_y(\tau)$ along the direction of the ribbon  (the period is $T$).
We choose the Coulomb gauge so that  the electric field is $E_y=-\frac{\partial A_y}{\partial \tau}$
(in the  ribbon the electric field is applied along the y-axis).
We relate the vector potential to the flux through
\begin{eqnarray}
\phi=A_y(\tau)a_c,
\label{flux}
\end{eqnarray}
where  $a_c$ is the unit cell length of the ribbon.
The magnitude of the vector potential is chosen so that during the time interval  $T$
the flux  $\phi/\phi_0$ changes  $2\pi$ (in units $\phi_0=\frac{\hbar c}{e}$).
The magnitude of the flux  $\phi$ will serve as an adiabatic  parameter of a cyclic adiabatic evolution.
Note that  particle-hole symmetry of the band structure is {\it intact} as the flux changes.

The polarization is obtained by adding of the Zak phases of the occupied subbands $l$
\begin{eqnarray}
\mathcal{P}=-e\sum_{l=1}^{l'}\int_{-\pi/a_c}^{\pi/a_c} \frac{dk}{2\pi} A^l_{k},\nonumber\\
\label{pol}
\end{eqnarray}
where $k$ is the wavevector along the ribbon direction (the vector potential appears with it, i.e., $k+A_y$).   Note that the subband index $l$ is  discrete while in Eq.(\ref{Zak}) the corresponding variable $\theta$ is continuous.
Here  the Berry vector potential is
\begin{eqnarray}
A^l_{k}&=&i\left\langle u_{lk}\right|\frac{\partial}{\partial\phi}\left|   u_{lk} \right\rangle,
\label{vecp}
\end{eqnarray}
where the periodic function $u_{lk}(r)$ is defined by
the Bloch wavefunction  $\psi_{lk}(r)=e^{ikr}u_{lk}(r)$.  When  there  are band crossings, i.e.,  degeneracy points, the Berry phase must be computed carefully.

One can
compute numerically  the polarization $\mathcal{P}$ using the following
{\it gauge-invariant} method\cite{Van,Res}
\begin{eqnarray}
\mathcal{P}=\frac{e}{2\pi}\sum_{s=0}^{N_s-1} \frac{1}{N_s}\mathrm{Arg}\left[   \mathrm{det}  \langle \tilde{u}_{lk_s}| \tilde{u}_{l'k_{s+1}} \rangle   \right].\nonumber\\
\end{eqnarray}
The periodic function $\tilde{u}_{lk}(r)$ is defined by
 $u_{lk}(r)=\frac{1}{\sqrt{N_s}}\tilde{u}_{lk}(r)$.  Note that the normalization of $\tilde{u}_{lk}(r)$ is such that $ \langle \tilde{u}_{lk}| \tilde{u}_{lk} \rangle=\int_{cell}dr \tilde{u}_{lq}^{*}(r)\tilde{u}_{lq}(r)=1$.
Here  we have discretized the Brillouin zone into small $N_s$ intervals and   the discrete  wavevectors are $q_s$ with the index $s=0,...,N_s-1$ so that $k_0=-\pi/a_c$ and $k_{N_s}=\pi/a_c$.
The overlap $\left\langle \tilde{u}_{l k_{s}}\right|\left.\tilde{u}_{l'k_{s+1}}\right\rangle$ is between  valence  subband wavefunctions  at the adjacent wavevectors.
It is computed using the wavefunctions obtained from a  tight-binding Hamiltonian.
The angle $\mathrm{Arg}\left[   \mathrm{det}  \langle \tilde{u}_{lq_s}| \tilde{u}_{l'q_{s+1}} \rangle   \right]$ can be chosen  from the interval  $[0,2\pi]$
or  $[-\pi,\pi]$\cite{ARG}.  In the first case the computed value of the Zak phase is $\pi$ while in the second case it is $0$. We have verified this for both periodic armchair and zigzag ribbons.  In the presence of inversion symmetry this  uncertainty related to two possible   positions of the center of the Wannier functions\cite{Zak}.    This result holds both for armchair  and zigzag ribbons.


\begin{references}
\bibitem{ARG} Note that $Arg[\prod_i z_i]\neq \sum_i Arg[z_i]$ since  $\sum_i Arg[z_i]$ can be much larger than $2\pi$.
\end{references}
\end{document}